\documentclass[fleqn,twoside]{article}
\usepackage{espcrc2}
\usepackage{graphicx}
\usepackage{amsmath}

\newcommand{\cO}{{\cal O}}
\newcommand{\be}{\begin{equation}} 
\newcommand{\ee}{\end{equation}} 
\newcommand{\bea}{\begin{eqnarray}} 
\newcommand{\eea}{\end{eqnarray}} 

\title{Quenched QCD with fixed-point and chirally improved fermions}

\author{
Christof Gattringer\address[unir]{Institut f\"ur Theoretische Physik, Universit\"at
Regensburg, D-93040 Regensburg, Germany}, 
Meinulf G\"ockeler\addressmark[unir]$^,$\address[unil]{Institut f\"ur Theoretische Physik, Universit\"at
Leipzig, D-04109 Leipzig, Germany},
Peter Hasenfratz\address[unib]{Institut f\"ur Theoretische Physik,
Universit\"at Bern, CH-3012 Bern, Switzerland},
Simon Hauswirth\addressmark[unib], 
Kieran Holland\address[unisd]{Department of Physics, University of
California at San Diego, San Diego, USA},
Thomas J\"org\addressmark[unib], 
K.J.~Juge\addressmark[unib],
C.B.~Lang\address[unig]{Institut f\"ur Theoretische Physik, Universit\"at
Graz, A-8010 Graz, Austria}, 
Ferenc Niedermayer\addressmark[unib],
P.E.L.~Rakow\addressmark[unir], 
Stefan Schaefer\addressmark[unir] and 
Andreas Sch\"afer\addressmark[unir] 
[BGR (Bern-Graz-Regensburg) collaboration]\thanks{This contribution is based
on parallel talks and posters presented by C.\ Gattringer, S.\ Hauswirth, K.\ Holland,
K.J.\ Juge, C.B.\ Lang and S.\ Schaefer at LATTICE 2002.}
                }
\begin{document}

\begin{abstract}
In this contribution we present results from quenched QCD simulations
with the parameterized fixed-point (FP) and the chirally improved (CI)
Dirac operator. Both these operators are approximate solutions of the
Ginsparg-Wilson equation and have good chiral properties. We focus our
discussion on observables sensitive to chirality. In particular we
explore pion masses down to 210 MeV in light hadron spectroscopy,
quenched chiral logs, the pion decay constant  and the pion scattering
length. We discuss finite volume effects, scaling properties of the FP
and CI operators and performance issues in their numerical
implementation.   
\vspace{1pc}
\end{abstract}

\maketitle

\section{Introductory remarks}

Fixed point \cite{fp1,fp2,fp3,fp4} and chirally improved fer\-mi\-ons
\cite{ci1,ci2} provide an approach to chiral symmetry on the lattice
based on the Ginsparg-Wilson equation \cite{GiWi82}. In an actual
implementation both the FP and CI fermions give rise to an approximate
solution of this equation and it has to be tested how well a chiral
fermion is described by such an approximation. Here we report on our
results from quenched QCD calculations focussing in particular on 
observables sensitive to chiral symmetry (for a recent general review
of results with chiral actions see \cite{giusti}).

In another contribution to these proceedings \cite{bgrplen} we gave a
more general overview of our calculations and also a brief introduction
to the construction of the FP and CI operators ($D_{FP}$, $D_{CI}$).
Here we would like to deepen the discussion and present selected topics
in more detail. These topics include light hadron spectroscopy, the
quenched chiral logarithm, the pion decay constant and a preliminary
study of the $I=2$ pion scattering length. 

Both the FP and the CI operators make use of the full Clifford  algebra
and a large number of gauge paths to approximate a solution of the
Ginsparg-Wilson equation. They differ in the method for determining the
coefficients in front of the individual terms of the Dirac operator.
The FP operator is constructed from classical equations which determine
the fixed point of a renormalization group transformation in QCD. For
the CI operator the Ginsparg-Wilson equation is mapped onto a system of
coupled equations for the coefficients of the individual terms of the
Dirac operator and this system is then solved numerically. In our
implementation  of the two operators we restrict ourselves to terms
essentially on the hypercube only. For a more detailed discussion of
the operators see \cite{fp1,fp2,fp3,fp4,ci1,ci2,bgrplen}. 

\begin{table}[tb]
\renewcommand{\arraystretch}{1.2} 
\begin{tabular}{c|c|c|c|r} \hline\hline
$D$ & $N_s^3\times N_t$ & $a(r_0)$ &  \#conf & $m_{\rm
PS}/m_{\rm V} $  \\ \hline
FP & $16^3\times 32 $ & 0.16 fm  & 200 & 0.28--0.88 \\ 
FP &  $12^3\times 24 $ & 0.16 fm & 200 & 0.3--0.88 \\
ov3 & $12^3\times 24 $ & 0.16 fm & 100 & 0.21--0.88 \\
FP & $8^3\times 24 $ & 0.16 fm   & 200 & 0.3--0.88 \\
FP & $12^3\times 24 $ & 0.10 fm  & 200 & 0.34--0.89 \\
FP & $16^3\times 32 $ & 0.08 fm  & 100 & 0.36--0.89 \\ \hline
CI & $16^3\times 32 $ & 0.15 fm & 100 & 0.38--0.85 \\
CI &  $12^3\times 24 $ & 0.15 fm& 100 & 0.36--0.85 \\
CI & $8^3\times 24 $ & 0.15 fm  & 200 & 0.33--0.85 \\
CI & $16^3\times 32 $ & 0.10 fm & 100 & 0.33--0.92 \\
CI & $12^3\times 24 $ & 0.10 fm & 100 & 0.32--0.92 \\
CI & $16^3\times 32 $ & 0.08 fm & 100 & 0.40--0.95 \\ \hline
\end{tabular}
\caption{Statistics for the FP and CI Dirac operator together with the
lattice spacing measured with the Sommer parameter and the range of
pseudoscalar to vector mass ratios we worked at. \label{table:1}}
\end{table}

The parameters of our quenched QCD calculations were chosen similar for
the two operators such that a direct comparison of the results is
possible. For both operators we used lattices  of size $8^3\times 24$,
$12^3\times 24$,  $16^3\times  32$ with  lattice spacings  of
approximately $a=0.08$ fm, $0.10$ fm and $0.15$ fm. Thus, we were able
to study finite volume effects (constant $a$) as well as 
discretization effects (constant physical volume). The FP operator was
used together with quenched  configurations from a parameterized
perfect gauge action \cite{perfgauge} subsequently smoothened with
renormalization group inspired smearing \cite{fp3}. The latter can be
viewed as part of the parameterization of the FP operator.   We also
employed a version of the FP operator (ov3) augmented with  overlap
steps \cite{overlap}. It is constructed from the parameterized FP
operator with three terms of  the Legendre expansion of the overlap. 
For the chirally improved operator we generated the  quenched gauge
configurations with the L\"uscher-Weisz gauge action \cite{luweact} and
applied one step of HYP blocking \cite{hypblocking}  to improve the
properties of the Dirac operator. An overview of the statistics is
given in Table~\ref{table:1}.  There we also give the results for the
lattice spacing as determined from the Sommer parameter
\cite{perfgauge,sommerci} and the range of pseudoscalar to vector meson
mass ratios we have worked at. The smallest pion mass we have worked at
is 210 MeV for the FP operator (160 MeV with ov3) and 240 MeV for the
CI operator. We did not encounter problems with exceptional
configurations for the listed range of pion masses and we expect that
we can reach pion masses of 200 MeV for both operators without having
to go to exceedingly small lattice spacings.

Our hadron propagators were created by smeared sources. For the FP
operator we fixed the gauge configurations to Coulomb gauge and used
Gaussian sources. For the CI operator we employed Jacobi smearing
\cite{jacobi} and no gauge fixing was necessary. 

At every fermion offset of the hypercube a large number of gauge paths
contribute in our Dirac operators. By precalculating and storing some
basic gauge link products the construction of the corresponding sparse
matrix could be speeded up significantly. Every offset on the hypercube
is represented by a color-Dirac matrix whose size is $12 \times 12$
independently of the number of gauge paths to this offset. When
comparing the cost to e.g.~the Wilson operator (without clover term)
the number of floating point operations in the basic matrix-vector
multiplication is increased by a factor of $\approx 36$.   However,
since the bottleneck is the access to memory and we do more operations
with each SU(3) matrix we fetch from memory (all 16 entries in the
Dirac space are used, Wilson uses only 4) the actual number might be
quite different. Furthermore we observed \cite{fp3} that some numerical
operations such as the computation of low lying eigenvalues converge
considerably faster for our Dirac operators than for Wilson's
operator. 

Our calculations were done on the Hitachi SR8000 at the
Leibniz Rechenzentrum in Munich and the production runs were typically
performed on 8 to 16 nodes each consisting of 8 CPUs with shared
memory. The Dirac operator was distributed among the nodes and the
parallelization was done using MPI. 

\section{Pion masses}

\subsection{Different correlators for the pion}

\begin{figure}[p]
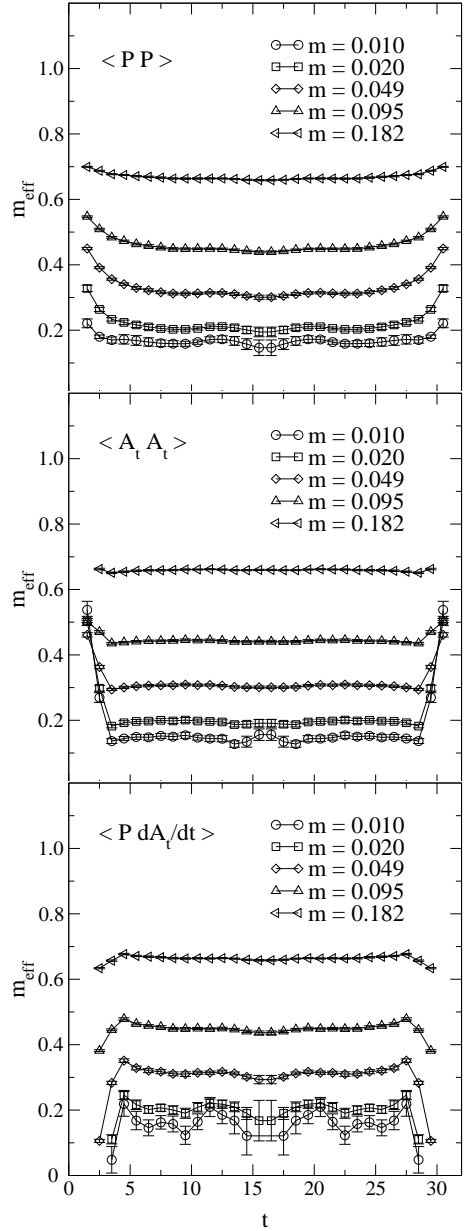

\hspace*{2mm}
\includegraphics[width=60mm,clip]{lat2002-twopt16x32b870_hyp_sp_2_meff.eps}
\vspace{-0.8cm}\\
\hspace*{2mm}
\includegraphics[width=60mm,clip]{lat2002-twopt16x32b870_hyp_sp_8_meff.eps}
\vspace{-0.8cm}\\
\hspace*{2mm}
\includegraphics[width=60mm,clip]{lat2002-twopt16x32b870_hyp_sp_12_meff.eps}
\caption{\label{pioncorrelators}
Effective mass plots for three different correlation functions 
(top to bottom: 
$\langle \,P\,P\,\rangle$,
$\langle \,A_t\, A_t\,\rangle$,
$\langle \,P\,\partial_t A_t\,\rangle$)
for $D_{CI}$, $16^3\times 32$, $\beta_{LW}=8.7$, $a=0.08~\textrm{fm}$; they
demonstrate the pion content of the involved operators.
As expected the signal becomes noisier from top to bottom.}
\end{figure}

In order to extract the pion we experimented with different operators
which have overlap with this channel. In Fig.~\ref{pioncorrelators} we
compare the effective mass plots at different bare quark masses
obtained for three combinations of sources and sinks constructed from
the point-like pseudoscalar and axial-vector densities. The mass
plateaus are quite well pronounced although for the third correlator
the signal is noisier as it is expected for a correlator with
additional derivatives. 

\begin{figure}[htb]
\includegraphics[width=70mm,clip]{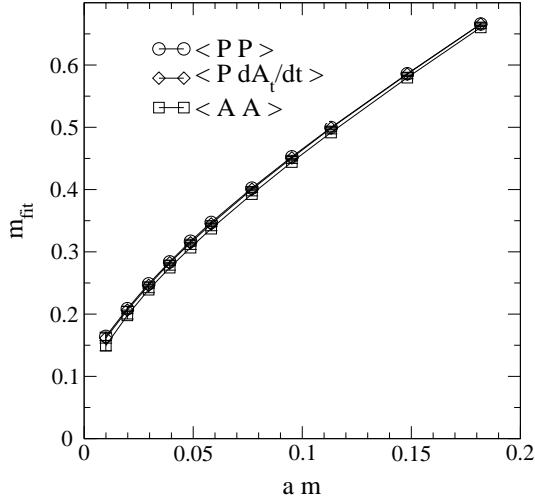}
\vspace{-4mm}
\caption{\label{fig:variouspions} 
Pion masses obtained from
different correlators 
($D_{CI}$, $16^3\times32$, $\beta_{LW}=8.7$, $a=0.08~\textrm{fm}$).}
\vspace{-4mm}
\end{figure}

The masses corresponding to the 2-point functions were extracted from
overall fits to cosh-behaviour covering  lattice distances bet\-ween
$(t_{\rm min},t_{\rm max})$. These parameters were fixed by inspecting
the effective mass plots and checking the $\chi^2$ of the fit.   In
Fig.~\ref{fig:variouspions} we compare the masses from the three
correlators and find that they agree within errors.

\subsection{Zero mode effects in pion correlators}

In the chiral region the pion correlators suffer from a topological
finite size effect specific to the quenched approximation
\cite{pioneffects,DoDrHo02}. Zero modes of the Dirac operator
(suppressed by the fermion determinant in a dynamical calculation) lead
to an unphysical  increase in the pion mass at small quark masses.
Since the abundance of zero modes scales  only as $\sqrt{V}$ with the
volume, the net effect goes as  $1/\sqrt{V}$. For small volumes,
however, this effect is quite important.

\begin{figure}[t!]
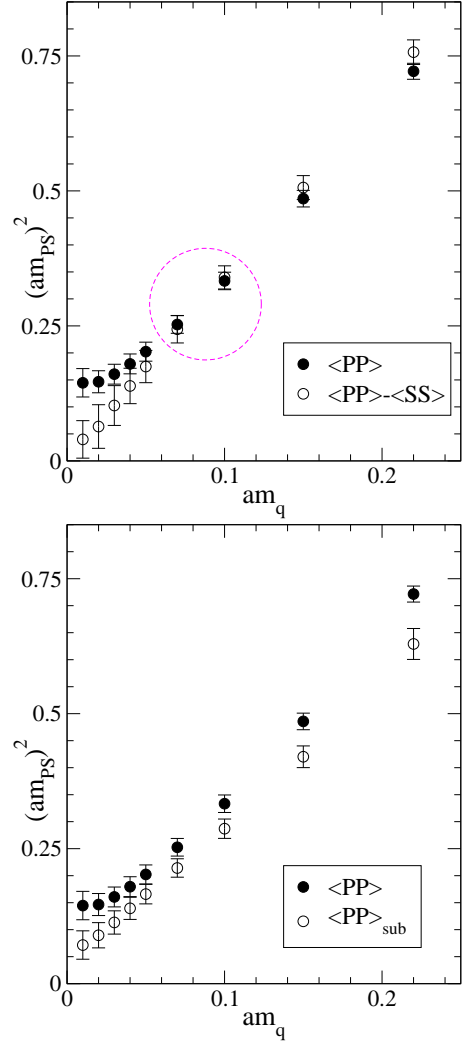

\hspace*{3mm}
\includegraphics[width=6cm]{zmcorr01.eps}\\
\hspace*{3mm}
\includegraphics[width=6cm]{zmcorr02.eps}
\vspace{-7mm}
\caption{Effect of the subtraction of the zero modes on the pion mass 
(FP-ov3 operator $6^3\times16$, $a=0.16~\mathrm{fm}$). The mass from
$G_{PP}$ is represented by filled circles, while we use open circles to
represent the results from the subtracted correlator of 
Eq.~(\protect{\ref{subcorr}}) (top plot), and from the
explicit removal of the zero modes from the quark propagators (bottom
plot). \label{fig:1}}
\vspace{-7mm}
\end{figure}

Removing simply the contribution of the zero modes from the quark
propagator is a dangerous procedure since it might change the pion
propagator in a non-local manner. Another possibility is
\cite{pioneffects,DoDrHo02} to consider the difference of the
pseudoscalar and scalar 2-point functions
\begin{equation}
G_{PP-SS}=G_{PP} -G_{SS} \ .
\label{subcorr}
\end{equation}
This method is based on the fact that for exactly chirally symmetric
actions the scalar propagator has the same topological finite size
effect as the pseudoscalar propagator, while for small quark masses 
the lightest particle  in this channel is much heavier than the pion.
This observation suggests the following strategy. At small quark
masses, where $G_{PP}$ is strongly distorted, we determine the pion
mass from $G_{PP-SS}$. Going towards heavier quark masses the
significance of the zero modes is suppressed and we expect a window
where $G_{PP}$ and $G_{PP-SS}$ lead to consistent mass fits. In this
window and beyond it we use the $G_{PP}$ correlator. At heavy quark
masses the mass difference bet\-ween the pseudoscalar and scalar masses
is too small to be resolved and the fitted mass from $G_{PP-SS}$ would
be larger than the true pseudoscalar mass. Fig.~\ref{fig:1} illustrates
these expectations on a small lattice for the FP-ov3 operator. In the
top plot the circle indicates the window where the zero mode effects
are already negligible, but the scalaris much heavier than the
pion  and so the correct pion mass is easily seen in  the $G_{PP-SS}$
correlator. The bottom figure illustrates the danger of separating
modes from the quark propagator itself: although at small quark masses
one obtains the same pion mass as from $G_{PP-SS}$, at intermediate and
large quark masses the pion mass obtained is incorrect. No window
exists in this case.

\begin{figure}[t]
\hspace*{-2mm}
\includegraphics[height=6.5cm,clip]{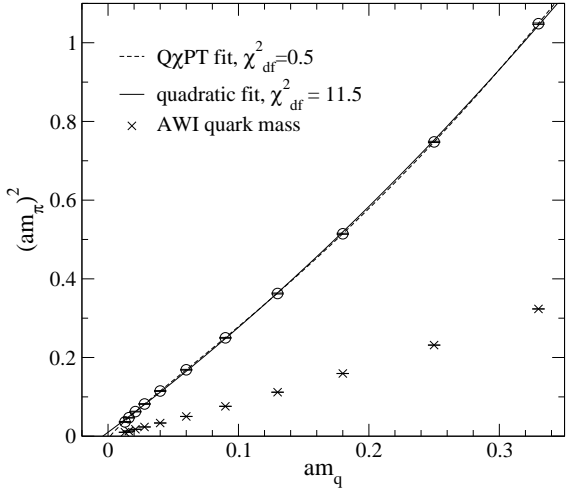}
\vspace{-7mm}
\caption{Pion mass (circles) and AWI mass (crosses) as a function of
the bare quark mass. $16^3 \times 32$
lattice, $a = 0.16$ fm, FP operator.}
\label{fig:mpivsmquark}
\vspace{-7mm}
\end{figure}

Once the lattice volume is sufficiently large the topological finite
size effect is no longer visible. In Fig.~\ref{fig:mpivsmquark} we show
the pion mass for a box with spatial extent of $L = 2.6$ fm ($16^3
\times 32$ lattice, $a = 0.16$ fm, FP operator). The data is fitted
with a quadratic polynomial (full curve) and also a curve including the
quenched chiral log \cite{quenchlog} (dashed curve). Although the
difference bet\-ween these fits is not seen on the scale of the figure,
the quality of the $Q \chi PT$ fit is significantly better. Both curves
extrapolate reasonably well to zero with the quark mass  indicating
that the topological finite size effect is negligible in this large
box. In principle one can determine the quenched chiral log parameter
directly from the pion with degenerate quark masses (see
e.g.~\cite{quenchlogresults} for recent such  determinations with the
overlap operator) but in the next section we study the quenched
chiral logs using a more effective method. Fig.~\ref{fig:mpivsmquark}
also includes data for the axial Ward identity (AWI) mass (crosses)
which we will discuss later.

\subsection{Quenched chiral logs}

\begin{figure}[t]
\vspace{-1mm}
\includegraphics[height=6cm]{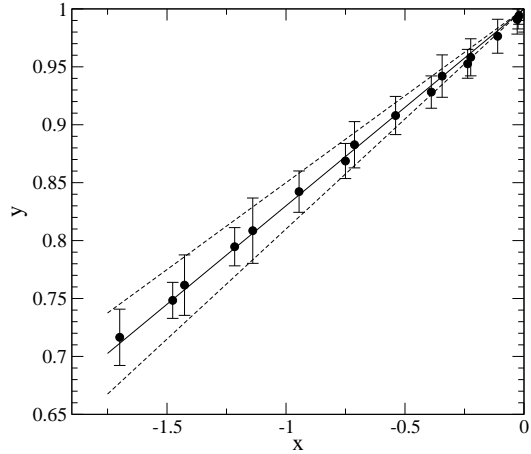}
\vspace{-5mm}
\caption{$x-y$ plot for the FP operator ($16^3\times32$ lattice, $a =
0.16$ fm). 
\label{fig:xyfp}}
\vspace{-5mm}
\end{figure}
\begin{figure}[t]
\includegraphics[height=6.3cm,clip]{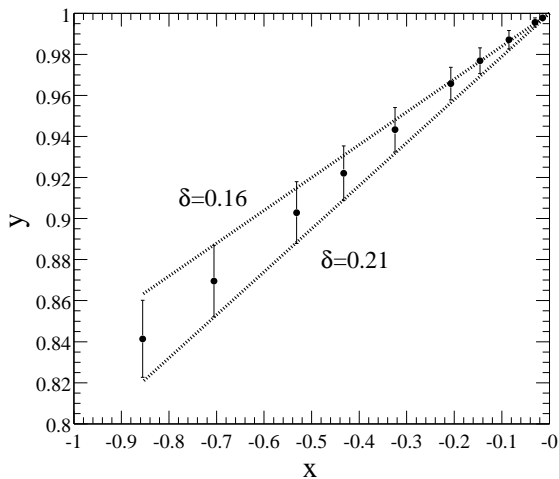}
\vspace{-7mm}
\caption{$x-y$ plot for the CI operator ($16^3\times32$ lattice, $a =
0.15$ fm).}
\label{fig:xyci}
\vspace{-5mm}
\end{figure}

Let us now discuss the determination of the quenched chiral log
parameter $\delta$ with a different method \cite{cppacsquench}.  The
idea is to use not only data for the pion with degenerate quark masses
but to also take into account data computed with non-degenerate quark
masses.  The prediction from quenched  chiral perturbation theory for
the dependence of the pseudoscalar meson mass $m_{\mathrm{PS},12}$  on
the quark masses $m_1$ and $m_2$ reads \cite{quenchlog}
\begin{equation}
\begin{split}
&m_{\mathrm{PS},12}^2= \\ &  A(m_1+m_2) \Big\{1-\delta
\big[\ln{\frac{2m_1A}{\Lambda_\chi^2}} +
\frac{m_2}{m_2-m_1}\ln{\frac{m_2}{m_1}} 
\big] \\
&+ \frac{1}{A}\alpha_X \big[m_1
\ln{\frac{2m_1A}{\Lambda_\chi^2}}+ m_2
\ln{\frac{2m_2A}{\Lambda_\chi^2}} \\ 
&+ \frac{m_1 m_2}{m_2\!-\!m_1}\ln \frac{m_2}{m_1}
\big]\Big\}+B(m_1\!+\!m_2)^2 +\cO(m^3, \delta^2),
\label{eq:ndmps}
\end{split}
\end{equation}
with $A$, $B$, $\delta$ and $\alpha_X$ a priori unknown constants. The
arbitrary scale $\Lambda_\chi$ is of the order $1~\mathrm{GeV}$.  The
dependence on the constants $A$, $B$ and $\Lambda_\chi$ can be  removed
by forming the following  cross ratio $y$ \cite{cppacsquench}:
\begin{equation}
y=
\frac{2 m_1}{m_1+m_2}\frac{m_{\mathrm{PS},12}^2}{m_{\mathrm{PS},11}^2}
\frac{2 m_2}{m_1+m_2}\frac{m_{\mathrm{PS},12}^2}{m_{\mathrm{PS},22}^2}\;.
\end{equation}
For small $\delta$, $\alpha_X$ and small quark masses $y$ is expected
to behave like
\begin{equation}
y=1+\delta x+\alpha_X z+\cO(m^2,\delta^2)
\label{eq:xyrel}
\end{equation}
with
\begin{eqnarray}
\!\!x&\! = \! &2+\frac{m_1+m_2}{m_1-m_2}\ln\left(\frac{m_2}{m_1}\right)
\;,\\
\!\!z&\! = \! &\frac{1}{A}\left(\frac{2 m_1 m_2}{m_2-m_1} 
\ln \frac{m_2}{m_1}\!-\!m_1-m_2
\right),
\end{eqnarray}
which means that $\delta$ can be extracted for small quark masses from
the slope of $y$ as a function of $x$. Since in this analysis the quark
mass enters only in ratios, for $m_1$ and $m_2$ any definition can be
used where the quark mass has no additive renormalization. (The
multiplicative quark mass renormalization cancels.) The mass $m_{AWI}$
defined by the axial Ward identity has this property and was used when
comparing Eq.~(\ref{eq:xyrel}) with the data.\footnote{The bare quark
mass in our Dirac operators has a small, but non-zero additive
renormalization, see Sect.3.1.} The plots are shown in
Fig.~\ref{fig:xyfp} for the FP operator ($16^3\times32$ lattice, $a =
0.16$ fm) and in Fig.~\ref{fig:xyci} for the CI operator
($16^3\times32$ lattice, $a = 0.15$ fm). Suppressing $\alpha_Xz$ and
other higher order terms in Eq.~(\ref{eq:xyrel}) we find
$\delta=0.17(2)$ for the FP operator and 
$\delta = 0.18(3)$ for the CI operator.

We have not yet studied the systematical errors of these predictions
due to cut-off or chiral symmetry breaking effects   (our Dirac
operators satisfy the Ginsparg-Wilson relation approximately only).
Further, the data with large negative $x$ come from large ratios of the
two quark masses which implies in our case that one of the quark masses
is large, stretching the validity of Eq.~(\ref{eq:ndmps}).

\section{Pion decay constant}

\subsection{The residual quark mass}

\begin{figure}[b!]
\includegraphics[width=70mm,clip]{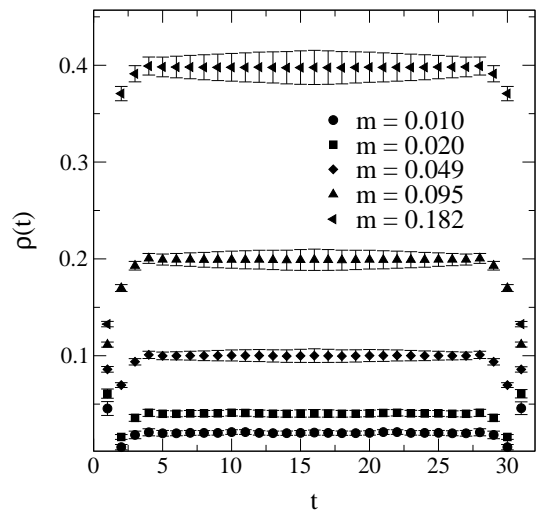}
\vspace{-5mm}
\caption{\label{fig:ratio_1}Results for the ratio $\rho(t)$ 
for $D_{CI}$, $16^3\times32$, $\beta_{LW}=8.7$, $a=0.08~\textrm{fm}$.}
\vspace{-5mm}
\end{figure}

For a chirally symmetric theory one may define conserved,
covariant\footnote{with respect to chiral transformations} currents
which do not require renormalization and covariant  scalar and
pseudoscalar densities whose renormalization factors satisfy $Z_S = Z_P
= 1/Z_m$ \cite{KiYa98c,fp4}. Here $Z_m$ is the renormalization factor
of the quark mass which is the coefficient of the covariant scalar
density in the Dirac operator. Although the overhead using such
currents and densities is expected to be relatively small \cite{fp4} we
used point-like, non-conserved and non-covariant operators in this
work. (The mass operator in the action is, however, covariant.) 

\begin{figure}[t!]
\includegraphics[width=70mm,clip]{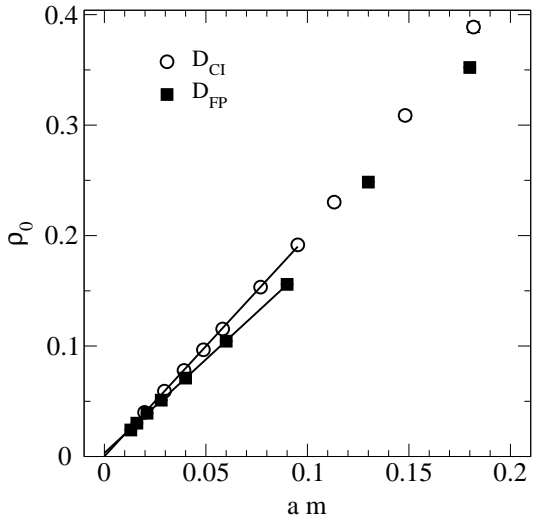}
\vspace{-5mm}
\caption{\label{fig:rho_0}The plateau values 
for the ratio $\rho_0(m)\equiv 2\,m_{AWI}(m)$
vs. the bare quark masses for both Dirac operators $D_{CI}$ 
($\beta_{LW}=7.9$, $a=0.15~\textrm{fm}$) and $D_{FP}$ ($\beta_{FP}=3.0$ , 
$a=0.16~\textrm{fm}$) for lattice
size $16^3 \times 32$. The linear fit indicates a very small 
residual quark mass.}
\end{figure}

In Fig.~\ref{fig:ratio_1} we show the ratio 
\be
\rho(t) = \frac{
  \langle P^-(0)\,\partial_t A^+_t(\vec{p}=0,t)\rangle 
}{\langle P^-(0)\,P^+(\vec{p}=0,t)\rangle }%
\label{eq:ratio}
\ee
where $P$ and $A_t$ are the pseudoscalar density and the time component
of the axial-vector current, respectively and $P^{\pm}=P^1 \pm i P^2$,
where 1,2 are flavor indices. This quantity exhibits excellent plateaus
for separations above 3 lattice spacings.  We denote the plateau values
by $\rho_0$. The sources $P(0)$ are smeared, the sink operators
unsmeared.

For renormalized fields the ratio is just twice the renormalized quark
mass (see also the discussion in the next section); in fact, $\rho_0/2$
may be employed to define a bare mass parameter. This mass has no
additive renormalization: $\rho_0/2=0$ if $m_\pi=0$. We have referred
to this mass before as the Ward identity mass $m_{AWI}$. We identify
the residual quark mass for our Dirac operators by the value of the
bare quark mass in the action for which $\rho_0/2=0$. 
Fig.~\ref{fig:rho_0} gives an example of our results, both for $D_{CI}$
and $D_{FP}$ (at slightly different lattice spacings) and we find
linear behaviour for small bare quark masses. A linear fit to these data
and data for other parameters demonstrates a very small residual quark
mass shift ranging for $D_{CI}$ from 0.002(1) at $a=0.15~\textrm{fm}$
to 0.000(1) at  $a=0.08~\textrm{fm}$, and for $D_{FP}$ from -0.0006(4) 
at $a=0.16~\textrm{fm}$ to -0.0194(2) at   $a=0.08~\textrm{fm}$. The
largest value for $D_{FP}$ is  likely related to the fact that $D_{FP}$
was optimized at $a=0.16~\textrm{fm}$ and then also used at the smaller
lattice spacing without readjustment.

\subsection{$Z_A$ and $f_\pi$}

Exact chiral symmetry on the lattice implies that the covariant
conserved axial current and the covariant pseudoscalar density satisfy
the equation
\be
\partial_\mu\, \tilde{A}^{\pm}_\mu(x)  = 2\,m\,\tilde{P}^{\pm}(x)\;,
\label{eq:ap}
\ee
which can be used in on-mass-shell Green's functions, or put 
differently, Eq.~(\ref{eq:ap}) is valid in general bare Green's
functions up to contact terms. In this equation $m$ is the bare quark
mass multiplying the covariant scalar density in the Dirac operator,
$\tilde{A}_\mu$ is the covariant conserved current and $\tilde{P}$ is
the covariant density. Eq.~(\ref{eq:ap}) is a consequence of bare Ward
identities which follow directly from the path integral formulation of
lattice QCD with Ginsparg-Wilson fermions. 

Considering Eq.~(\ref{eq:ratio}) with such covariant and conserved
operators we get $2m$ on the left hand side. Replacing the conserved
covariant current on the r.h.s. by $Z_A A_t$, where $A_t$ is the naive
point-like current, we can calculate $Z_A$ in terms of the correlators
and $m$. It can be shown \cite{fp4} that, for $2R=1$ in the
Ginsparg-Wilson relation as it is the case for $D_{CI}$,  in these
correlators the covariant pseudoscalar density can be replaced by the
naive point-like density divided by $(1-m/2)(1-m^2/4)$.  This factor
goes to 1 in the chiral or in the continuum limit. At the end the
renormalization factor $Z_A$ of the point-like current can be obtained
by measuring correlators of the point-like operators. Up to factors
($\approx 1$) mentioned above  $Z_A=2m/\rho_0$, where $\rho_0$ is the
plateau of the ratio in Eq.~(\ref{eq:ratio}) discussed in the previous
section. We collected the corresponding numbers in
Table~\ref{table:ZA}. These numbers might be distorted by the small
chiral symmetry breaking effects in $D_{CI}$. The trick of replacing
the covariant densities by their naive point-like versions is modified
for the case when $2R$ is a non-trivial local operator \cite{fp4} as it
is the case for $D_{FP}$. For the FP action the corresponding
correlators are not yet measured.

\begin{table}[t]
\begin{tabular}{@{}lrlll}
   \multicolumn{5}{c}{$D_{CI}$} \\
\hline
$N_s^3\times N_t$ &$a=$ & $0.15~\textrm{fm}$ & 
$0.10~\textrm{fm}$ & $0.08~\textrm{fm}$\\
\hline
$ 8^3 \times 24$ && 0.94(2) &        &         \\
$12^3\times 24$ && 1.00(2) & 0.96(1)&         \\
$16^3\times 32$ && 1.00(2) & 0.97(1)& 0.96(1) \\
\hline
\end{tabular}\\[2pt]
\caption{Values for the renormalization factor $Z_A$ of the point-like
axial current for $D_{CI}$. 
\label{table:ZA}}
\end{table}

Similar considerations can be used to connect $f_\pi$ with the
correlator of point-like, naive pseudoscalar densities. Start with the
correlator  $\sum_{\vec x}\langle \tilde{P}(\vec{x} ,t)
\tilde{P}(0)\rangle$ where $\tilde{P}$ is the covariant density
entering the AWI. Saturating this correlator by the pion intermediate
state for large time separation, using Eq.~(\ref{eq:ap}) and the
definition of the pion decay constant\footnote{In our convention the
experimental value of $f_\pi$ is $\approx 131$MeV.} one obtains
\be\label{eq:fpi}
\begin{array}{l}
\sum_{\vec x}\langle \tilde{P}^-(\vec{x} ,t) \tilde{P}^+(0)\rangle\cr
\displaystyle\quad\quad = - \frac{f_\pi^2 m_\pi^3}{8m^2} \left(e^{-m_\pi\,t}+
e^{-m_\pi\,(T-t)} \right)\;.
\end{array}
\ee
As before, the covariant pseudoscalar density in Eq.~(\ref{eq:fpi}) can
be replaced by the point-like density divided by $(1-m/2)(1-m^2/4)$ in
the case of the CI action. 

The considerations of this section wer based on the assumption of exact
chiral symmetry. When using these results to interpret our data, we
assumed that the chiral symmetry breaking effects in the CI action were
small enough to be neglected. However, this assumption needs to be
checked explicitly.

\begin{figure}[t]
\includegraphics[width=70mm,clip]{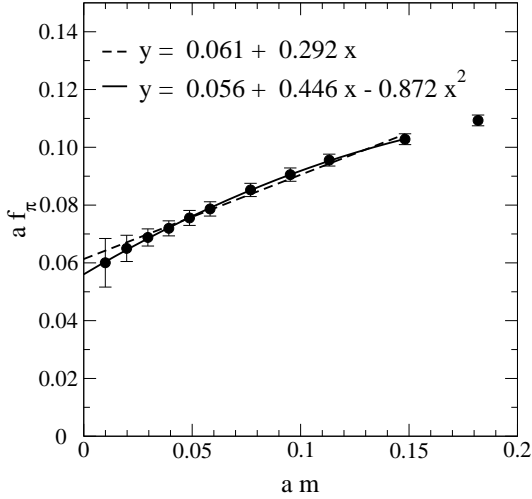}
\vspace{-7mm}
\caption{\label{fig:fpifit}Results for $f_\pi$
for $D_{CI}$, $16^3 \times 32$, $\beta_{LW}=8.7$, $a=0.08~\textrm{fm}$ 
together with fits linear or quadratic in $a\,m$.}
\vspace{-5mm}
\end{figure}

Since the source is typically smeared, we measured propagators from the
smeared source to point sinks and smeared sinks. Appropriate ratios
then allow us to recover e.g. unsmeared-unsmeared type propagators.

Fig.~\ref{fig:fpifit} shows $f_\pi$ as a function of the bare mass  for
$D_{CI}$ comparing a linear and a  quadratic fit.  For the
extrapolation to the chiral limit we use the results from the quadratic
fit.  The chiral extrapolations of $a f_\pi$ in Figs. 10 and 11   show
excellent scaling behaviour.

\begin{figure}[t]
\includegraphics[width=70mm,clip]{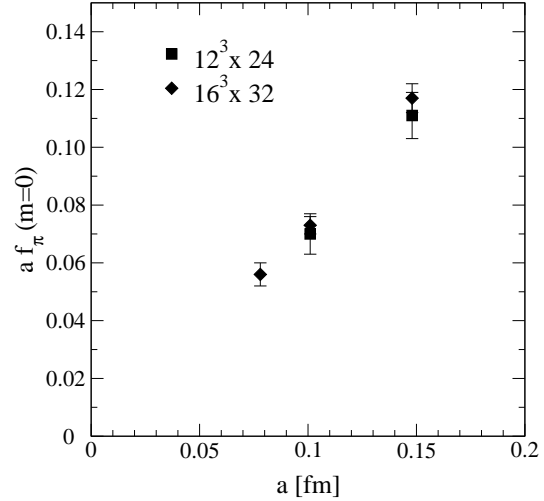}
\vspace{-5mm}
\caption{\label{fig:fpi_scaling_1}The (quadratic) extrapolations of
$f_\pi$ to the chiral limit for different lattice spacings exhibit
simple scaling behaviour (results for $D_{CI}$).}
\end{figure}

\begin{figure}[htb]
\includegraphics[width=70mm,clip]{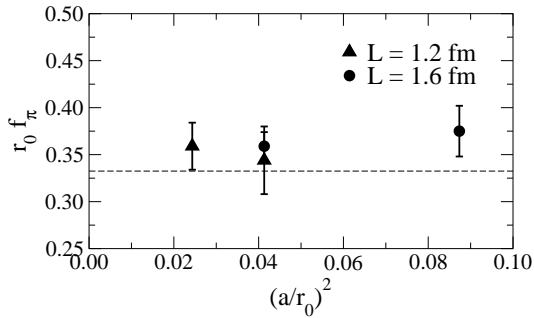}
\caption{\label{fig:fpi_scaling_2}The (quadratic) extrapolations of
$f_\pi$ to the chiral limit for different lattice spacings in
dimensionless units ($r_0$ denotes the Sommer parameter) exhibit simple
scaling behaviour (results for $D_{CI}$). The dashed line corresponds to
the  experimental (i.e. unquenched) value 131 MeV.}
\end{figure}

\section{Light hadron spectroscopy}

The parameters of our simulations (Table~\ref{table:1}) allow us to
make a scaling study with both actions in a fixed volume $L \approx
1.3$~fm at lattice constants $a \approx 0.08$, $0.10$ and $0.15$~fm and
a finite volume analysis at fixed $a \approx 0.15$~fm in boxes with
$L=1.3-2.6$~fm. We considered $\sim 10$ different quark masses and
inverted the Dirac matrix with a multimass solver. Note that  the
hadron masses at different quark masses with other parameters fixed are
strongly correlated.

With this setup we could study different aspects of light hadron
spectroscopy close to the chiral limit. The subject is old,
nevertheless we found surprises, results which we do not quite
understand and which require further study.

\subsection{Hadron spectroscopy in terms of hadronic scales}

Since our main interest lies in understanding the behaviour of the Dirac
operators, it is natural to fix the scale within the hadronic sector
itself. In Sect. 4.3 we shall connect these results to the
typical gauge sector scale $r_0$ also.

An obvious candidate to carry the scale could be the vector meson mass
$m_V$ at the quark mass where $m_{PS}/m_V$ is equal to the
experimental value. Identifying this vector meson mass with $m_\rho
\approx 770$ MeV gives the cutoff and all the masses in MeV. This
choice for the scale has the disadvantage that it refers to the rho
mass deep in the chiral limit and is plagued by relatively large
statistical errors. For many purposes (like scaling and finite volume
studies) it is better to fix the scale by the vector meson mass 
$m_V(x)$, where $m_{PS}/m_V=x$ fixed and chosen in a range
where $m_V(x)$ can be precisely determined. In our ana\-ly\-sis we
have taken $x=0.75$. A similar method was used earlier in the works
\cite{from_giusti}. 

\begin{figure*}[h]
\centering
\includegraphics[height=85mm,width=140mm]{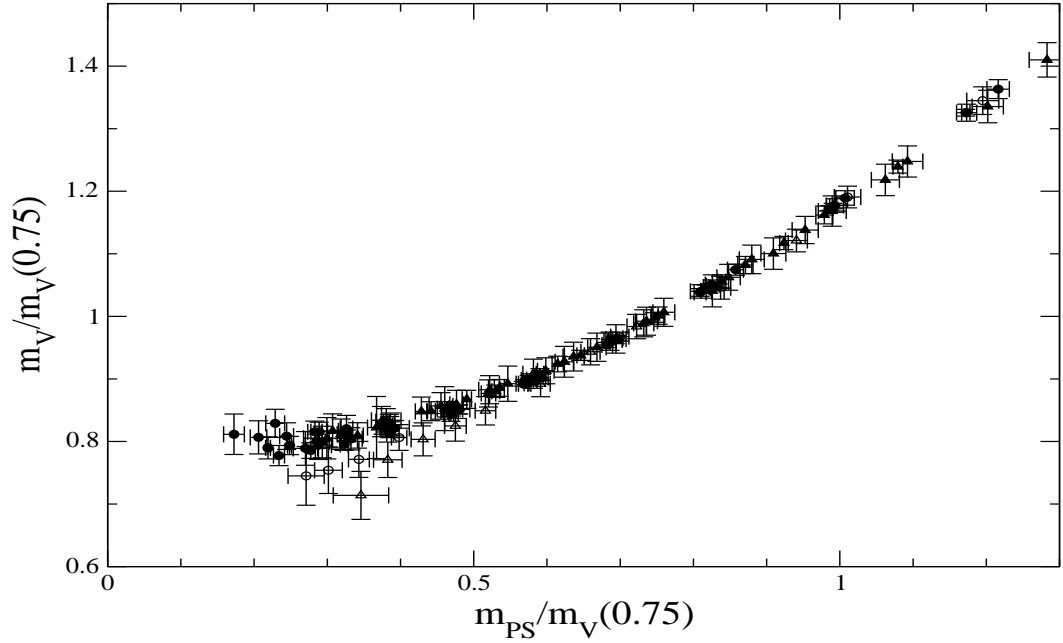}
\vspace*{-8mm}
\caption{\label{fig:rho} 
All the results on the vector and pseudoscalar masses
of our simulations with
different lattice spacings, volumes and actions. 
The scale is carried by the vector meson mass where $m_{PS}/m_V=0.75$.} 
\end{figure*}

Fig.~\ref{fig:rho} gives the vector meson mass $m_V$ as a function of the
pseudoscalar mass $m_{PS}$ in $m_V(0.75)$ units. All our simulation results
from Table~\ref{table:1} are collected in this figure\footnote{Here and in the
following, for the clarity of the figure, we suppress points with very large
errors.}. Independently of the lattice scale $a$, the size of the box
($L=1.3-2.6$~fm) and for both Dirac operators the points lie on a universal
curve. This observation goes beyond our work: collecting data from other large
scale simulations performed with different Dirac operators (Wilson, clover,
staggered fermions with or without improvement) the results lie on this curve
\cite{next_paper}. The point (0.75,1) of this curve is fixed by definition; the
other points seem to be then independent of the cut-off in the range of recent
simulations and in volumes down to $L \approx 1.3$~fm.

\begin{figure*}[h]
\centering
\includegraphics[height=85mm,width=140mm]{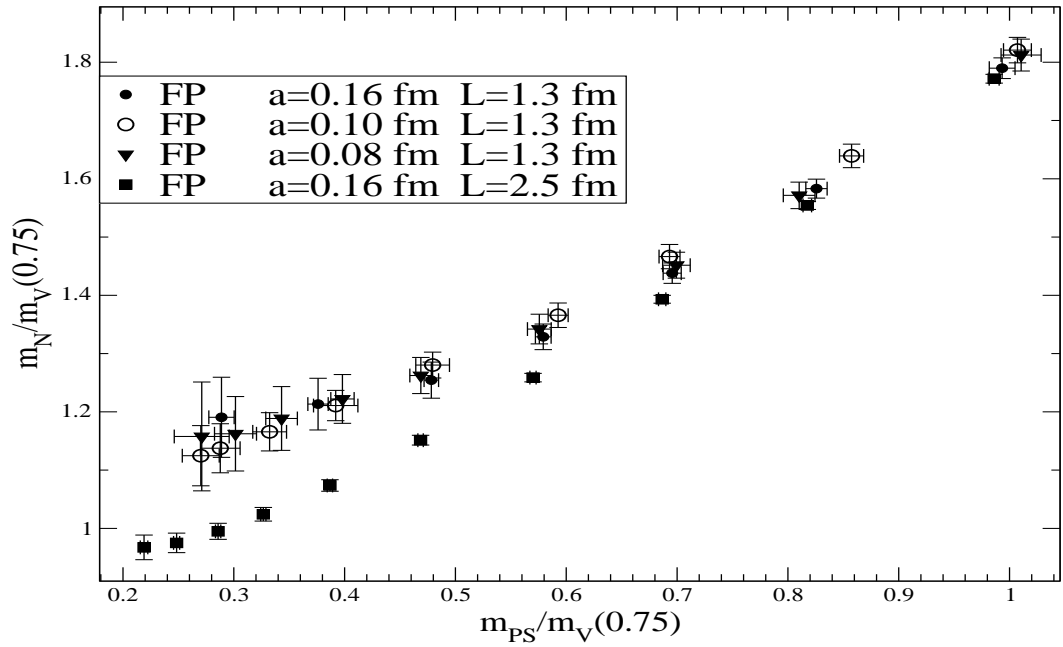}
\vspace*{-8mm}
\caption{\label{fig:nucleon} 
The nucleon mass as a function of the pseudoscalar mass. The upper
curve is a scaling test in a fixed volume.}
\end{figure*}

Fig.~\ref{fig:nucleon} shows the nucleon mass as a function of $m_{PS}$ in
$m_V(0.75)$ units for the FP Dirac operator.  The points forming the upper
curve all refer to   $L \approx 1.3$~fm, they represent a scaling test for $a
\approx (0.08-0.16)$~fm. Within the statistical errors no scaling violation is
seen in this figure. The conclusion is the same for the CI operator. The large
volume ($L\approx 2.5$ fm) results form the lower curve. These points were
obtained on a coarse lattice with $a \approx 0.16$~fm. 

For heavy quark masses we expect that the nucleon finite size effects will
become smaller. We see indeed in Fig.~\ref{fig:nucleon} that the two curves join for
heavy quarks. This has an additional message. Since the upper curve scales, we
might assume it is close to the continuum limit. This suggests that the large
volume curve (for which we do not have a scaling test) is also, at least for
heavy quarks, close to the continuum limit.

\begin{figure*}[h]
\centering
\includegraphics[height=85mm,width=140mm]{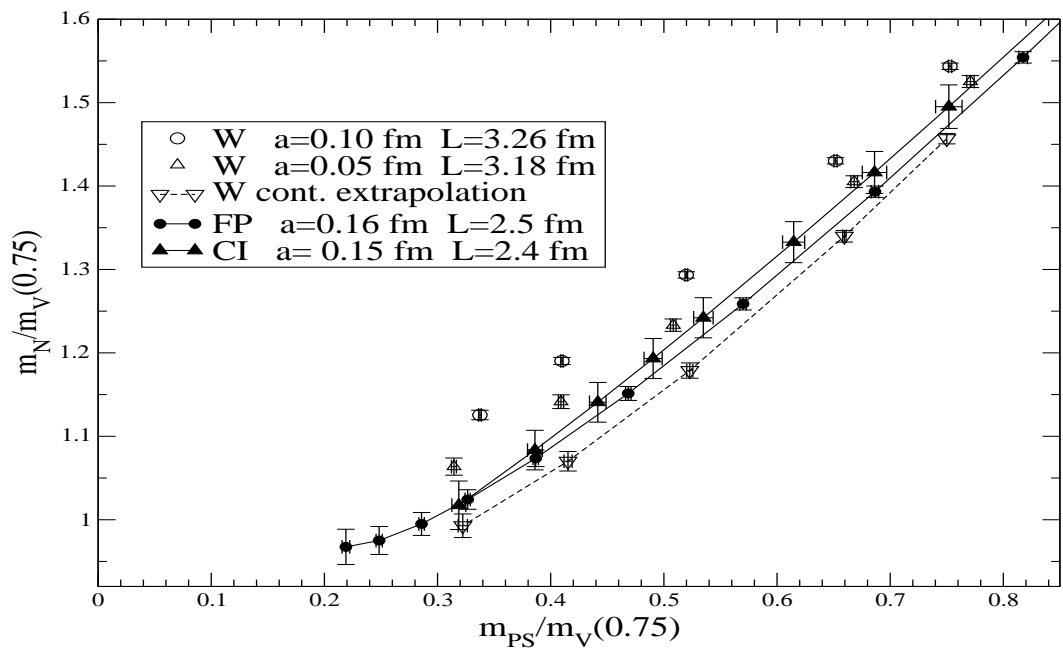}
\vspace*{-8mm}
\caption{\label{fig:nucleon_large} 
Similar to Fig.~\ref{fig:nucleon}. The large volume FP and CI results obtained on coarse configurations are compared with Wilson data and their continuum limit extrapolation \cite{cppacsquench}.}
\end{figure*}

Fig.~\ref{fig:nucleon_large} illustrates that the scaling of the
$D_{FP}$ data in Fig.~\ref{fig:nucleon} is not 'built in'
in the quantity studied. In this figure our large volume results
are compared with the CP-PACS data from their
recently completed analysis \cite{cppacsquench}. These results were
obtained with the Wilson Dirac operator at four different lattice
spacings ($a=0.05-0.10$~fm). For the clarity of the figure we plotted
the $a=0.05$~fm and $a=0.10$~fm data only. Even the $a=0.05$~fm points
are rather far from the continuum as estimated in
\cite{cppacsquench}. The difficulty of such a continuum
extrapolation is illustrated in  Fig.~\ref{fig:CP-PACS_cont}. 

\clearpage

\clearpage

\begin{figure*}[h]
\centering
\includegraphics[height=85mm,width=140mm]{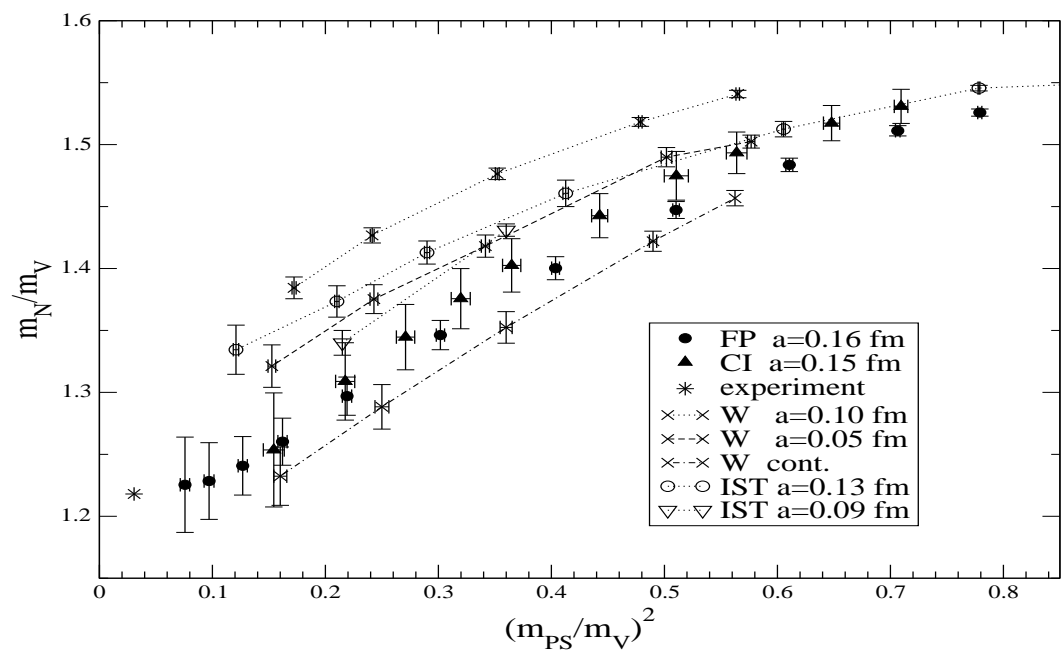}
\vspace*{-8mm}
\caption{\label{fig:ape} 
APE plot where the FP and CI results are compared with Wilson \cite{cppacsquench} and improved staggered data \cite{MILC13}.}
\end{figure*}

\clearpage

The four sets of points in Fig. \ref{fig:CP-PACS_cont} refer to $x=m_{PS}/m_V=$ 0.75, 0.7, 0.6 and 0.4. For each set a linear extrapolation is done using the results at 4 different lattice spacings. The data and the extrapolated points at $a=0$ are taken from \cite{cppacsquench}. The 4 points on the r.h.s. of Fig.~\ref{fig:CP-PACS_cont} are the FP numbers at $a \approx 0.16$~fm for the 4 different $x$-values. Further results are needed to make a firm conclusion from Fig.~\ref{fig:nucleon_large}.

\begin{figure}[tbp]
\centering
\includegraphics[height=70mm,width=73mm]{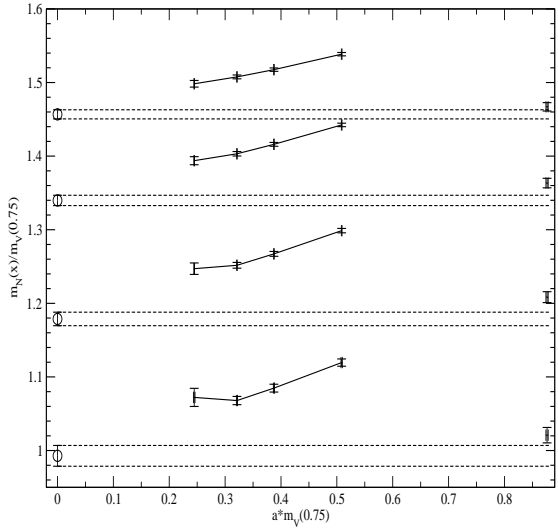}
\vspace{-8mm}
\caption{\label{fig:CP-PACS_cont} 
The continuum extrapolation of the CP-PACS Wilson data \cite{cppacsquench} for 4 different x-values. The points on the far right are the FP $a\approx0.16$ fm results.}
\end{figure}

Fig.~\ref{fig:ape} is a standard APE plot where the large volume, coarse
lattice FP $a\approx 0.16$~fm and CI  $a\approx 0.15$~fm data  are compared
with the  CP-PACS results 
\cite{cppacsquench} and with the improved staggered fermion data obtained at
$a=0.13$~fm \cite{MILC13} and $a=0.09$~fm \cite{MILC09} (2 points only). The
coarse FP and CI data are significantly closer to the CP-PACS continuum
extrapolated curve than the fine lattice Wilson or the improved staggered
points. On the other hand, as we remarked above, the long continuum
extrapolation of the Wilson data \cite{cppacsquench} is not an easy task.

\subsection{Dispersion relation and the speed of light}

Another quantity which can be studied within the hadronic sector is the
energy-momentum dispersion relation. As earlier investigations show,
$E=E({\vec p}\,)$ is a cut-off sensitive quantity and the deviations
from the continuum form $(E^2({\vec p}\,)-m^2)/{\vec p}^{\;2}= c^2$ can
be large. Here $c$, the speed of light, should be independent of  ${\vec
p}$ and equal to 1 in our convention. For the FP Dirac operator at $a
\approx 0.16$~fm in a box with $L \approx 2.5$~fm the energy $E({\vec
p}\,)$ could be determined with relatively small errors in the momentum
range $|{\vec p}\,| = (0.5-1.0)$~GeV and the corresponding dispersion
relation can not be distinguished from its continuum form 
\cite{fp3,next_paper}. 

\begin{figure}[tbp]
\includegraphics[width=73mm]{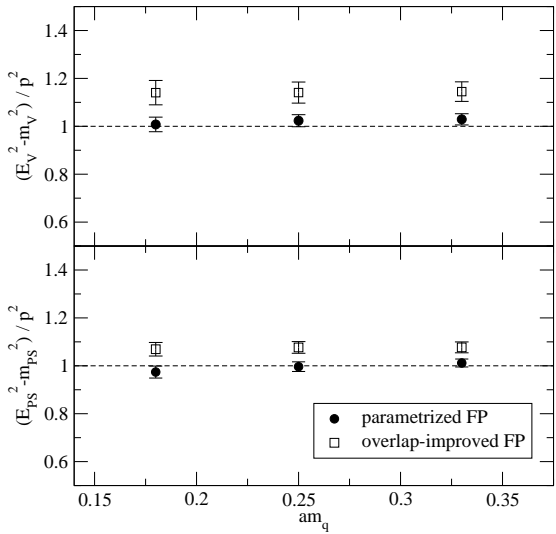}
\caption{\label{fig:speed} 
The speed of light as obtained from the vector and pseudoscalar channels as a function of the quark mass. }
\end{figure}

Fig.~\ref{fig:speed} shows the speed of light as a function of the quark
mass for the vector and pseudoscalar meson as measured by the FP and the
3 overlap steps augmented FP Dirac operators. For both hadrons the FP
operator results show no deviation from 1, while the data with the
overlap augmented operator lie above 1 beyond the statistical errors. We
find this result difficult to understand. Starting with the
parameterized FP operator which approximates quite well a GW solution,
we expected to produce small changes only by performing 3 overlap
steps\footnote{It is true, however, that in the spectrum of the overlap
operator with the parametrized FP kernel even the low lying eigenvalues
have a small shift along the circle relative to the spectrum of the
kernel, see Fig(5.4) in the second reference in \cite{fp3}.}. 

\begin{figure}[h]
\vspace{-0.8mm}
\includegraphics[width=73mm,height=71mm]{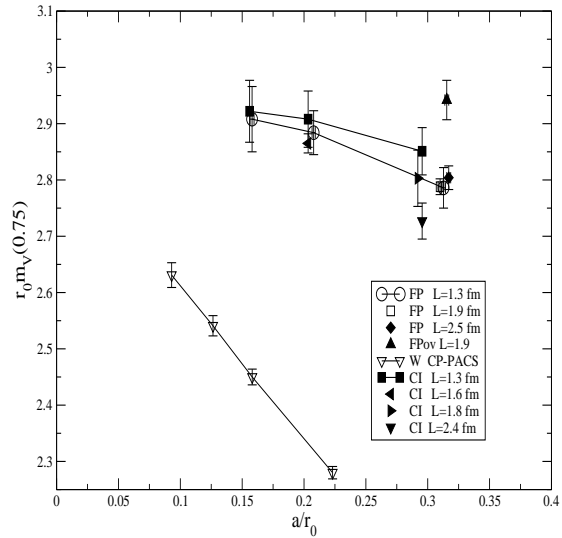}
\vspace{-8mm}
\caption{\label{fig:r0scale} 
Scaling test for $r_0m_V(0.75)$. }
\end{figure}

\subsection{Connecting the hadronic and gauge scales} 

In order to connect the hadron sector with the gauge sector it is
sufficient to relate the reference scale $m_V(0.75)$ we used above with
a gauge theory scale, like $r_0$ \cite{sommer}. The Sommer parameter
has been measured for both actions close to the $\beta$- range of our
simulations \cite{perfgauge,sommerci}. 

Fig.~\ref{fig:r0scale} shows $r_0 m_V(0.75)$ as a function of the
lattice resolution. The 3 FP data on the coarsest lattice with
different volumes $L=1.3-2.5$~fm completely coincide showing that
$m_V(0.75)$ has no finite size effect within the small errors in these
boxes. This is consistent with Fig.~\ref{fig:rho}. The 3 CI data on the
coarsest lattice are consistent with the FP number, but have larger
errors.

Relative to this coarsest point the $a\approx$0.10 and 0.08~fm data lie
higher and indicate cut-off effects in $r_0 m_V(0.75)$. We have to
remember though that measuring $r_0$ on a coarse lattice is a
difficult, perhaps not even quite well defined problem.

As Fig.~\ref{fig:r0scale} shows, the prediction of the overlap
augmented FP Dirac operator on the coarsest lattice is definitely
inconsistent with that of the FP result. Since the Dirac operators FP
and ov3 were used on the same gauge configurations (therefore they see
the same $a$ and $r_0$ in a quenched theory) we have to conclude that
one or both Dirac operators have clearly identified cut-off effects in
$m_V(0.75)$ at $a \approx 0.16$~fm. For comparison we also included the
Wilson data \cite{cppacsquench} in Fig.\ref{fig:r0scale}.

\section{Pion scattering length}

The pion scattering length is an important observable characterizing
dynamical effects of the strong interaction and its ab initio
calculation on the lattice is an important nonperturbative test of QCD.
In the literature several attempts to compute the scattering length 
with Wilson fermions \cite{Sharpe,Fukugita,CPPACS,Liu,JLQCD} and
staggered fermions \cite{Sharpe,Fukugita} can be found. However, 
calculations with a chiral Dirac operator which allows for a better
control of the small mass region are still missing. In full QCD the
scattering length is a quantity which vanishes in the chiral limit, 
while it is power divergent in the quenched theory 
\cite{Bernard} and eventually a study in the full theory is
desirable. This is the initial stage of our investigation which we 
regard as a preparation for a simulation in the full theory when 
unquenched configurations become available. In this preliminary study
we reuse the FP propagators computed for the spectroscopy to calculate
the $\pi\pi$ S-wave scattering length in the $I=2$ channel.
Although using these propagators with only a single source limits the 
quality of the overlap with the pion scattering state the results are quite encouraging and a more refined investigation is in preparation.

\subsection{Method}

In order to calculate the $\pi^+\pi^+$ scattering length from a 
Euclidean lattice simulation, we use L\"uscher's relation
\cite{Luscher} that relates the energy shift of the two pion state,
$\Delta E(L)$, in a finite volume ($V=L^3$) to the scattering length,
$a_0$, in the infinite volume limit to order $1/L^5$, 
\begin{equation}
\Delta E=-\frac{4\pi a_0}{m_\pi L^3}\left\{1+c_1\frac{a_0}{L}+
c_2\frac{a_0^2}{L^2}+\cdots\right\} \;,
\end{equation}
where $c_1 = -8.9136$ and $c_2 = 62.9205$ are numerical constants
computed in \cite{Luscher}. Propagators generated for the
spectroscopy  study were used to construct the two pion state and to
measure its energy,  $E=2\,m_\pi+\Delta E$. In other words, a single
Gaussian source was  used to create the two pions on the gauge fixed
configurations.  We also applied the same Gaussian smearing at the
sinks in this study.  We worked on the $8^3 \times 24$ and $12^3
\times 24$ lattices at $a = 0.16$ fm. All the scattering lengths
shown in Fig.~\ref{summary} are  for $m_\pi\,L>4$ except for the two
lightest masses for $L=12\,a$  where $m_\pi\,L$ is only larger than
$3$. 

\subsection{Analysis}

We extract the energy shift of the two pion state in a finite volume 
by considering ratios of two pion correlation functions and single
pion correlation functions. We follow the procedure and  notation of
Ref.~\cite{Sharpe} and construct the ``direct'' (D) and  ``crossed''
(C) contractions of
\begin{equation}
\left\langle\sum_{\vec{x}_1}{\mathcal O}_1(\vec{x}_1,t)
\sum_{\vec{x}_2}{\mathcal O}_2(\vec{x}_2,t){\mathcal S}(t=0){\mathcal
S}(t=0)\right\rangle \;,
\end{equation}
where ${\mathcal O}$ annihilates a $\pi^+$ and ${\mathcal S}$ are the
gaussian sources for the pions. Note that D and C actually refer to
the ratio after dividing by two single pion correlators. In practice
we fit the numerator and denominator separately. The different
diagrams (contractions) of ratios are labeled,
\\
\\
$D(t)\sim$ glue exchange between the two pions\\
$C(t)\sim$ quark exchange between the two pions
\\
\\
where the combination of interest ($I=2$) is $D(t)-C(t)$ whose large 
$t$ behaviour is approximately $Ze^{-\Delta Et}$. We have also
expanded  the exponential to leading order in $\Delta Et$ in the fit,
but there were no differences within the errors. 

Periodic boundary conditions were used in the time direction so that 
we must take into account a backwards propagating pion. The fitting
form of the two-particle correlator (the numerator of $D(t)-C(t)$) is
\begin{equation}
Z_{\pi\pi}(e^{-(2m_\pi+\Delta E) t}+
e^{-(2m_\pi+\Delta E)(T-t)}+ze^{-m_\pi T}) \; .
\end{equation}
We simultaneously fit the pion propagator entering the denominator for
$Z_\pi$ and $m_\pi$,
\begin{equation}
Z_{\pi}(e^{-m_\pi t}+e^{-m_\pi(T-t)}) \; ,
\end{equation}
resulting in the following five fitting parameters:  $m_\pi, Z_\pi,
\Delta E, Z_{\pi\pi}$ and $z$. We perform a fully correlated fit to
the two correlation functions. In the two figures that follow, we
show the fits for the following set of parameters,\\
\centerline{$\beta_{FP}=3.00$, $a \approx 0.16$ fm}
\vskip 0.5mm
\centerline{$N_s=12$, $N_t=T=24$}
\vskip 0.5mm
\centerline{ $m_\pi/m_\rho=0.41$, $200$ configurations}
\vskip 2mm
\noindent which is representative of all the other fits.

\begin{figure}[tbp]
\vspace*{-7mm}
\includegraphics[width=77mm,clip]{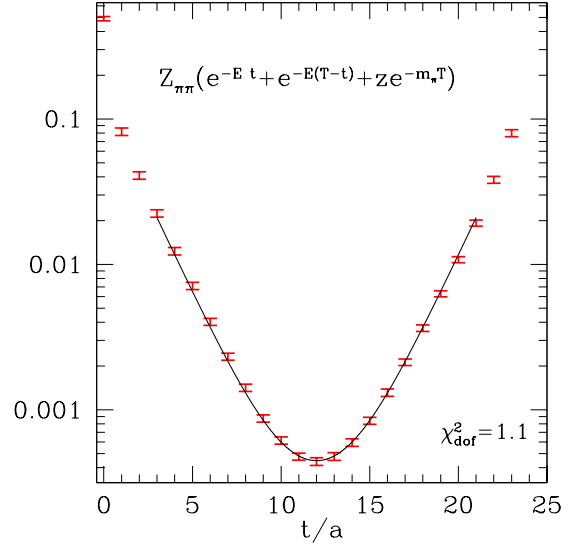}
\vspace{-5mm}
\caption{The numerator of the $D-C$ function.}
\label{DC}
\vspace{-6mm}
\end{figure}

\begin{figure}[tbp]
\hspace*{-3mm}
\includegraphics[width=77mm,clip]{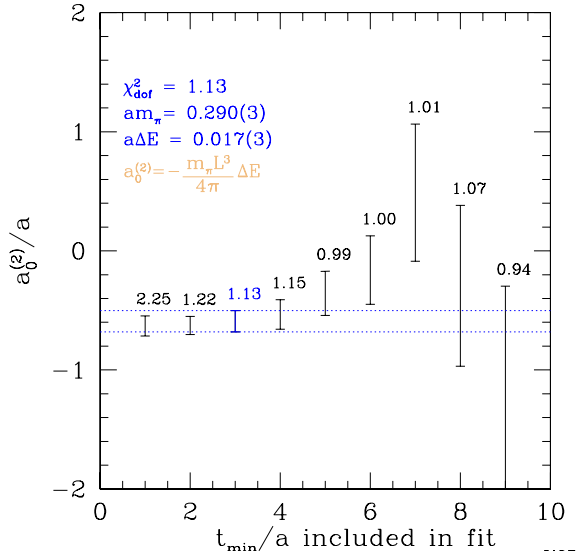}
\vspace{-5mm}
\caption{Fit results for $a_0$ vs $t_{min}$. 
The dashed lines mark the value chosen.}
\label{a0tmin}
\vspace{-5mm}
\end{figure}

\begin{figure}[ht]
\hspace*{-3mm}
\includegraphics[width=77mm,clip]{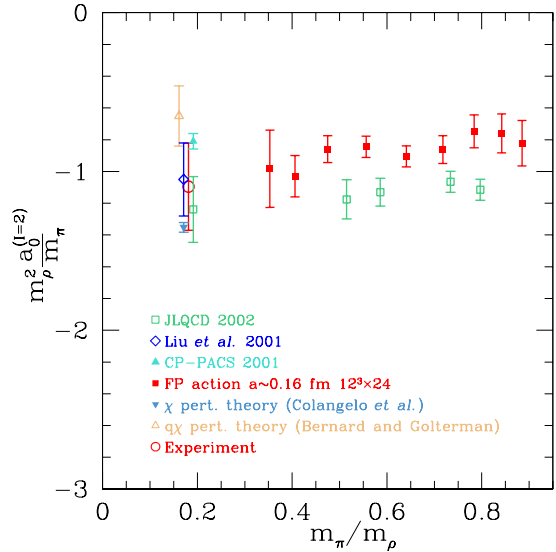}
\vspace{-5mm}
\caption{Summary of our results and other recent calculations. } 
\label{summary}
\vspace{-5mm}
\end{figure}

In Fig.~\ref{a0tmin}, we show how the extracted $a_0$ depends on the
initial time slice included in the fit. We see that nearly any time
slice from $t=2\,a$ gives the same results (with reasonable $\chi^2$
values) within the errors. This may be somewhat surprising
considering that our two pion sources are on top of each other. We
take the value for the fit starting at $t_{min}=3\,a$ in this
particular case. There were substantial difficulties in getting
results for the other lattice spacings as the temporal extent of the
box was not as long and/or short of statistics. In the following, we
mainly discuss the results from the lattice described above. 

\subsection{Results and discussion}

Using the parameterized fixed-point action, we have extracted the
$\pi^+\pi^+$ S-wave scattering length at threshold with the
propagators generated for spectroscopy studies. We were able to
obtain results for the $\beta_{FP}=3.0, 12^3\times24$ lattice and for
some of the larger masses on the $8^3\times 24$ lattice. In
Fig.~\ref{summary}, we plot the dimensionless quantity (as was used
in Ref.~\cite{Liu}), $m_\rho^2 a_0/m_\pi$, against $m_\pi/m_\rho$. We
also include in this plot some recent results from other groups as
well as quenched chiral perturbation theory \cite{Bernard} and chiral
perturbation theory predictions \cite{Colangelo}. Our raw results on
a coarse lattice are in reasonable agreement with Wilson/clover
action results on finer lattices or continuum limit extrapolated
numbers. Since we mostly only have results from a single coupling, no
real scaling tests could be performed for $a_0$. 

However, agreement between two of the coarser lattices in
limited cases for $m_\pi L>4$ and $m_\pi/m_\rho$ ratios larger than
$0.6$ indicate that they are small (also found in  spectroscopy
studies). The difficulties of extracting the numbers from the other
lattices are likely due to contamination from higher momentum pion
states. These effects can be reduced by introducing another source,
which is  currently under study.  As remarked above some zero mode
effects were observed in the pseudoscalar propagator in the
spectroscopy study for $m_{PS}/m_V$ mass ratios less than $0.5$.
These finite size effects as well as other systematic error studies
are work in progress.

\section{Summary and conclusions}

We have reported here several calculations of quantities relevant for
light quark phenomenology using two different Dirac operators which
are approximate solutions of the Ginsparg-Wilson equation. 

Both actions, beyond having good chiral behaviour, exhibited 
significantly smaller cut-off effects at $a\sim0.15$~fm than the
standard Wilson action at $a\sim0.05$~fm and the improved-staggered
action at $a\sim0.09$~fm on hadronic quantities that we have
surveyed.

We believe that the improvement more than compensates the apparent
increase in the cost of simulating our operators and that a study of 
dynamical simulations with these actions is a relevant challenge for
the future. 

\vskip5mm
\noindent
{\bf Acknowledgements:} 
The calculations were done on the Hitachi SR8000 at the Leibniz
Rechenzentrum in Munich and at the Swiss Center for Scientific
Computing in Manno. We thank the LRZ staff for training and support. 
This work was supported in parts by DFG,  BMBF and SNF and by  the
European Community's Human Potential Programme under contract
HPRN-CT-2000-00145. C.\ Gattringer acknowledges support by the
Austrian Academy of Sciences (APART 654).


\begin{thebibliography}{9}

\bibitem{fp1}
P.\ Hasenfratz, F.\ Niedermayer, Nucl.\ Phys.\ B 414 (1994) 785.

\bibitem{fp2} 
P.\ Hasenfratz, S.\ Hauswirth, K.\ Holland, T.\ J\"org, F.\
Niedermayer 
and U.\ Wenger,
Int.\ J.\ Mod.\ Phys.\ C 12 (2001) 691.

\bibitem{fp3} 
S.\ Hauswirth, {\sl Light hadron spectroscopy in quenched lattice 
QCD with chiral fixed-point fermions}, Thesis, Bern University 2002,
hep-lat/0204015;
T.\ J\"org, {\sl Chiral measurements in quenched lattice QCD with 
fixed-point fermions}, Thesis, Bern University 2002,
hep-lat/0206025.

\bibitem{fp4}
P.\ Hasenfratz, S.\ Hauswirth, T.\ J\"org, F.\ Niedermayer and K.\ Holland,
hep-lat/0205010.

\bibitem{ci1}
C.~Gattringer, Phys.~Rev.~D 63 (2001) 114501.

\bibitem{ci2}
C.\ Gattringer, I.\ Hip, C.B.\ Lang, Nucl.\ Phys.\ B 597 (2001) 451.

\bibitem{GiWi82} P.~Ginsparg, K.G.~Wilson, 
Phys.\ Rev.\ D 25 (1982) 2649.

\bibitem{giusti}
L.~Giusti, {\sl Exact chiral symmetry on the lattice: QCD
applications}, plenary talk at Lattice' 02, these proceedings.

\bibitem{bgrplen} C.~Gattringer, these proceedings, hep-lat/0208056.

\bibitem{perfgauge}
F.~Niedermayer, P.~R\"ufenacht and U.~Wenger, Nucl.\ Phys.\ B 597 (2001) 413.

\bibitem{overlap} 
R.\ Narayanan and H.\ Neuberger, Phys.\ Lett.\ B 302 (1993) 62,
Nucl.\ Phys.\ B 443 (1995) 305.

\bibitem{luweact}
M.\ L{\"u}scher and P.\ Weisz, Commun.\ Math.\ Phys.\ 97 (1985) 59;
Err.: 98 (1985) 433;
G.\ Curci, P.\ Menotti and G.\ Paffuti, Phys.\ Lett.\ B 130 (1983) 205,
Err.: B 135 (1984) 516.

\bibitem{hypblocking} 
A.~Hasenfratz and F.~Knechtli,
Phys.~Rev.~D 64 (2001) 034504.

\bibitem{sommerci}
C.\ Gattringer, R.\ Hoffmann and S.\ Schaefer,
Phys.\ Rev.\ D 65 (2002) 094503.

\bibitem{jacobi}
C.\ Best, M.\ G\"ockeler, R.\ Horsley, E.M.\ Ilgenfritz, H.\ Perlt,
P.E.L.\ Rakow, A.\ Sch\"afer, G.\ Schierholz, A.\ Schiller and S.\ Schramm
Phys.\ Rev.\ D 56 (1997) 2743;

C.\ R.\ Allton {\it et al.} (UKQCD Collaboration),
Phys.\ Rev.\ D 47 (1993) 5128.

\bibitem{pioneffects}
T.~Blum {\it et al.},
hep-lat/0007038;

\bibitem{DoDrHo02}
S.J.\ Dong, T.\ Draper, I.\ Horv\'ath, F.X.\ Lee, K.F.\ Liu and J.B.\ Zhang,
Phys.\ Rev.\ D 65 (2002) 054507.

\bibitem{quenchlog}
S.R.~Sharpe, Phys.~Rev.~D 46 (1992) 3146;
C.W.\ Bernard and M.F.L.\ Golterman, Phys.\ Rev.\ D 46 (1992) 853.   

\bibitem{quenchlogresults}
T.~W.~Chiu and T.~H.~Hsieh,
Phys.\ Rev.\ D 66 (2002) 014506;
T.\ Draper, S.J.\ Dong, I.\ Horv\'ath, F.X.\ Lee, K.F.\ Liu, N.\
Mathur and J.B.\ Zhang, hep-lat/0208045.

\bibitem{from_giusti}
C. R. Allton {\it et al.}, Nucl.\ Phys.\ B 489 (1997) 427; 
L. Giusti, C. Hoelbling and C. Rebbi, Phys.~Rev.~D 64 (2001) 11450.

\bibitem{cppacsquench}
S.~Aoki {\it et al.} (CP-PACS collaboration), 
hep-lat/0206009.

\bibitem{sommer}
R.~Sommer, Nucl.\ Phys.\ B 411 (1994) 839.
ALPHA, M.~Guagnelli, R.~Sommer and H.~Wittig, 
Nucl.\ Phys.\ B 535 (1998) 389.

\bibitem{GiHoRe01}
L. Giusti, C. Hoelbling, and C. Rebbi,
\newblock Phys. Rev. D 64 (2001) 114508;
Err. ibid. D65 (2002) 079903.

\bibitem{KiYa98c}
Y. Kikukawa and A. Yamada,
\newblock hep-lat/9810024.

\bibitem{MILC09}
D. Toussaint {\it et al.}, Nucl.\ Phys. B (Proc.\ Suppl.) 106 (2002) 111.

\bibitem{MILC13}
C. Bernard {\it et al.}, Phys.\ Rev.\ D 64 (2001) 054506.

\bibitem{next_paper}
BGR Collaboration, in preparation.

\bibitem{Sharpe}
  S.R.~Sharpe, R.~Gupta and G.W.~Kilcup, Nucl.\ Phys.\ B 383 (1992) 309.
 
\bibitem{Fukugita}
  M.~Fukugita, Y.~Kuramashi, M.~Okawa, H.~Mino and A.~Ukawa, 
   Phys.\ Rev.\ D 52 (1995) 3003.

\bibitem{CPPACS}
   CP-PACS Collaboration, Nucl.\ Phys. B (Proc.\ Suppl.) 106 (2002) 230.

\bibitem{Liu}
  C.~Liu, J.~Zhang, Y.~Chen and J.P.~Ma, Nucl.\ Phys.\ B 624 (2002) 360.

\bibitem{JLQCD}
   JLQCD Collaboration, hep-lat/0206011.

\bibitem{Bernard}
   C.~Bernard and M.~Golterman, Phys.\ Rev.\ D 53 (1996) 476.

\bibitem{Luscher}
   M.~L\"uscher, Nucl.\ Phys.\ B 354 (1991) 531.

\bibitem{Colangelo}
   G.~Colangelo, J.~Gasser and H.~Leutwyler, Nucl.\ Phys.\ B 603 (2001) 125.

\end{thebibliography}
\end{document}